\documentclass[review]{elsarticle}
\usepackage{lineno,hyperref}
\usepackage[english]{babel}
\usepackage[numbers]{natbib}
\usepackage{xcolor}
\usepackage{latexsym,amsmath,amssymb,amsbsy,graphicx,geometry}
\modulolinenumbers[5]

\begin{document}
	
\begin{frontmatter}
			
\title{Porous amorphous nitinol synthesized by argon injection: a molecular dynamics study}

\author[kfu]{A.A. Tsygankov\corref{cor1}}
\cortext[cor1]{Corresponding author}
\ead{tsigankov.artiom@gmail.com}

\author[kfu,ufrc]{B.N. Galimzyanov}
\ead{bulatgnmail@gmail.com}

\author[kfu,ufrc]{A.V. Mokshin}
\ead{anatolii.mokshin@mail.ru}

\address[kfu]{Kazan Federal University, 420008 Kazan, Russia} 
\address[ufrc]{Udmurt Federal Research Center of the Ural Branch of the Russian Academy of Sciences, 426067 Izhevsk, Russia}

\begin{abstract}
Porous crystalline nitinol is widely applied in various fields of science and technology due to the unique combination of physical and mechanical properties as well as biocompatibility. Porous amorphous nitinol is characterized by improved mechanical properties compared to its crystalline analogues. Moreover, this material is more promising from the point of view of fundamental study and practical application. The production of porous amorphous nitinol is a difficult task requiring rapid cooling protocol and optimal conditions to form a stable porous structure. In the present work, based on the results of molecular dynamics simulations, we show that porous nitinol with the amorphous matrix can be obtained by injection of argon into a liquid melt followed by rapid cooling of the resulting mixture. We find that the porosity of the system increases exponentially with increasing fraction of injected argon. It has been established that the system should contain about $\sim18$--$23$\% argon for obtain an open porous structure, while the system is destroyed by overheated inert gas when the argon fraction is more than $\sim23$\%. It is shown that the method of argon injection makes it possible to obtain a highly porous system with the porosity $\sim70$\% consisting the spongy porous structure similar to aerogels and metallic foams.
\end{abstract}
	

\end{frontmatter}

\section{Introduction}
Porous titanium nickelide alloy Ni$_{50}$Ti$_{50}$ known as nitinol is promising due to unique combination of physical and mechanical properties such as open porosity, biocompatibility, shape memory effect and high strength~\cite{nitinol_main,review_niti_applictions}. Therefore, porous nitinol is preferable in medicine, aerospace and transport industries for production of construction details, implants and filter elements~\cite{review_niti2,Bansiddhi_2011}. The presence of such properties is largely determined by the crystal structure that supports the austenitic-martensitic (B2$\rightarrow$B19') and martensitic-austenitic (B19'$\rightarrow$B2) transitions under thermal and mechanical influences~\cite{nitinol_main}. The main advantage of porous crystalline nitinol is the simplicity of its preparation and the availability of initial powder raw materials for synthesis, for example, by powder metallurgy methods~\cite{niti_production}. The synthesized porous samples usually have an open porous structure with the average linear size of the pores $\sim 200$\,$\mu$m, that is close to the pore size in bone tissues~\cite{Augat_2006}. At the same time, porous crystalline nitinol has some limitations, which include increased fragility (with increasing porosity, the fragility also become larger) and the tendency to form cracks~\cite{niti_lifetime,nakas2014_fatigue}.

From point of view of fundamental studies and practical applications, porous amorphous nitinol is of great interest~\cite{nitinol_amorph_rem}. This material is less exposed to crack formation due to the absence of defects, which inevitably occur in the case of the crystalline analogue. It has an increased resistance to external deformations that is confirmed by the results of molecular dynamics simulations. For example, it was shown in Ref.~\cite{GBN_2021,Panico_2008_test_porous_niti} that the Young's modulus of porous amorphous nitinol at tensile and compression deformations is more than twice larger then in the case of the crystalline material with the porosity $\phi\in[7.5;\,35]$\%. Despite these advantages, it is difficult to synthesize experimentally porous amorphous nitinol. The main reason for this is the need to apply the quench with extremely high rates (more than $1\times10^{6}$\,K/s) to obtain a stable amorphous structure~\cite{niti_isobaric}. Difficulties may also arise at choice an appropriate method for the synthesis of a porous structure. The known methods for preparing porous alloys, such as powder sintering~\cite{Anikeev_2020_2d_powder_niti}, additive technologies~\cite{Karaji_2006_3d_print}, spark plasma sintering~\cite{Bansiddhi_2011}, and self-propagating high-temperature synthesis~\cite{Shiskovsky_2012_sls_porous_nitinol,Biswas_2005_SHS}, can lead to formation of a crystalline structure in the powder contact zones~\cite{Galimzyanov_Yarullin_2018,Galimzyanov_Nikiforov_2020}. The presence of crystalline inclusions can lead to the degradation in the mechanical characteristics of the porous alloy, for example, due to an increase in the probability of the nanosized cracks formation at the interface between crystal and amorphous phases~\cite{JIANG_2013_amorph_cryst,Galimzyanov_Mokshin_2018}. Therefore, the method of direct melt foaming by gas injection is the most preferable for obtaining the amorphous structure without crystalline inclusions~\cite{Banhart_2001,Brothers_2006}. This method is widely used to form the pores in various metal melts by gas injection (most often by argon) under high pressure or gas-evolving foaming agents (for example, calcium oxide CaO, titanium hydride TiH$_{2}$) or by means of obtaining oversaturated metal-gas solutions followed by cooling the foamed liquid melt. However, this method is difficult to apply for production of porous amorphous nitinol due to the high melting point of Ni$_{50}$Ti$_{50}$ alloy ($\sim1570$\,K) and due to the lack of information about the optimal conditions for foaming the corresponding melt. Insufficient understanding of the pore formation mechanisms in the nitinol melt requires detailed studies related to molecular dynamics simulations.

In the present work, the process of liquid nitinol foaming via argon injection is studied by means of molecular dynamics simulations. The simulation conditions are close to the experimental ones implemented, for example, in the method of direct melt foaming by gas injection~\cite{Banhart_2001,Brothers_2006}. We determine the optimal fraction of the injected argon and the thermodynamic parameters of the system to obtain the porous amorphous alloy with the required porosity and the required pore morphology. The possibility of obtaining porous amorphous nitinol with closed pores (porosity up to $35$\%) and with an open percolating porous structure (with porosity up to $70$\%) is demonstrated. In Section $2$, the melt foaming procedure and the details of the applied hybrid interparticle interaction potential are discussed. Section $3$ is devoted to discussion of the obtained results. The conclusion is given in Section $4$.

\section{Argon injection procedure}

Initially, crystalline nitinol consisting $68\,750$ atoms of Ni and $68\,750$ atoms of Ti was chosen. Molecular dynamics simulations are performed in the isobaric-isothermal (NPT) ensemble with the time step $1$\,fs using the Lammps simulation package~\cite{lammps}. At all stages of the simulations, the pressure is equal to $1$\,atm.

The interaction between Ni, Ti and Ar atoms is specified using the hybrid interaction potential. For Ni-Ni, Ti-Ti and Ni-Ti interactions, the 2NN MEAM potential is applied to compute the total energy of the system~\cite{niti_pot}:
\begin{equation}\label{eq_2nnmeam_pot}
E = \sum_{i=1}^{N}\left[F_{i}(\rho_{i})+\frac{1}{2}\sum_{j\neq i}^{N}S_{ij}\phi_{ij}(r_{ij})\right].  
\end{equation}
Here, $F_i$ is the embedding function for the atom $i$ within a background electron density $\rho_i$; the pair potential $\phi_{ij}(r_{ij})$ and screening function $S_{ij}$ are evaluated at the distance $r_{ij}$ between atoms $i$ and $j$. In the 2NN MEAM potential, the pairwise interaction $\phi_{ij}(r_{ij})$ is not assigned a simple functional expression. As a rule, the value of the quantity $\phi_{ij}(r_{ij})$ is estimated by the embedding energy and the energy per atom
\begin{equation}\label{eq_2nnmeam_pair}
\phi_{ij}(r_{ij}) = \frac{2}{Z_{1}}\left\{E_{i}^{u}(r_{ij})-F_{i}(\rho_{i})\right\}.  
\end{equation}
In Eq. (\ref{eq_2nnmeam_pair}), $Z_1$ is the number of nearest-neighbor atoms, $E_{i}^{u}(r_{ij})$ is the energy per atom~\cite{Lee_Baskes_2000}:
\begin{equation}\label{eq_2nnmeam_Eu}
E_{i}^{u}(r_{ij}) = -E_{c}(1+a^{*}+da^{*3})\mathrm{e}^{-a^{*}},  
\end{equation}
where
\begin{equation}\label{eq_2nnmeam_a}
a^{*} = \sqrt{\frac{9B\Omega}{E_{c}}}\left(\frac{r_{ij}}{r_{e}}-1\right).  
\end{equation}
Here, $r_e$ is the equilibrium nearest-neighbor distance; $E_c$ is the cohesive energy; $d$ is an adjustable parameter; $B$ is the bulk modulus; $\Omega$ is the equilibrium atomic volume of the reference structure. The values of these parameters for Ni-Ti system are given in Table IV of Ref.~\cite{niti_pot}. The cutoff radius of the 2NN MEAM potential is $5.0$~\AA~for considered binary system. For Ni-Ar, Ti-Ar and Ar-Ar interactions, the binary Lennard-Jones potential is applied~\cite{lj_proof}: 
\begin{equation}\label{eq_LJ}
E_{LJ}(r_{ij}) = 4\epsilon_{\alpha\beta}\left[\left(\frac{\sigma_{\alpha\beta}}{r_{ij}}\right)^{12}-\left(\frac{\sigma_{\alpha\beta}}{r_{ij}}\right)^{6}\right],\,\,\alpha,\beta=\{\mathrm{Ni, Ti, Ar}\}.  
\end{equation}
The parameters of this potential -- the potential well depth $\epsilon_{\alpha\beta}$ and the effective atom diameter $\sigma_{\alpha\beta}$ -- are given as follows: $\epsilon_{\mathrm{ArTi}} =8.7\cdot10^{-2}$~eV and $\sigma_{\mathrm{ArTi}} =3$~\AA~for Ar-Ti interaction; $\epsilon_{\mathrm{ArNi}} =7.9\cdot10^{-2}$~eV and $\sigma_{\mathrm{ArNi}} =3.04$~\AA~for Ar-Ni; $\epsilon_{\mathrm{ArAr}} =1\cdot10^{-2}$~eV and $\sigma_{\mathrm{ArAr}} =3.41$~\AA~for Ar-Ar~\cite{Hansen_1969_lj_pot}. The applied cutoff radius of the Lennard-Jones potential is $2.5\sigma_{\alpha\beta}$.

\begin{figure}[ht!]
	\centering
	\includegraphics[width=1.0\linewidth]{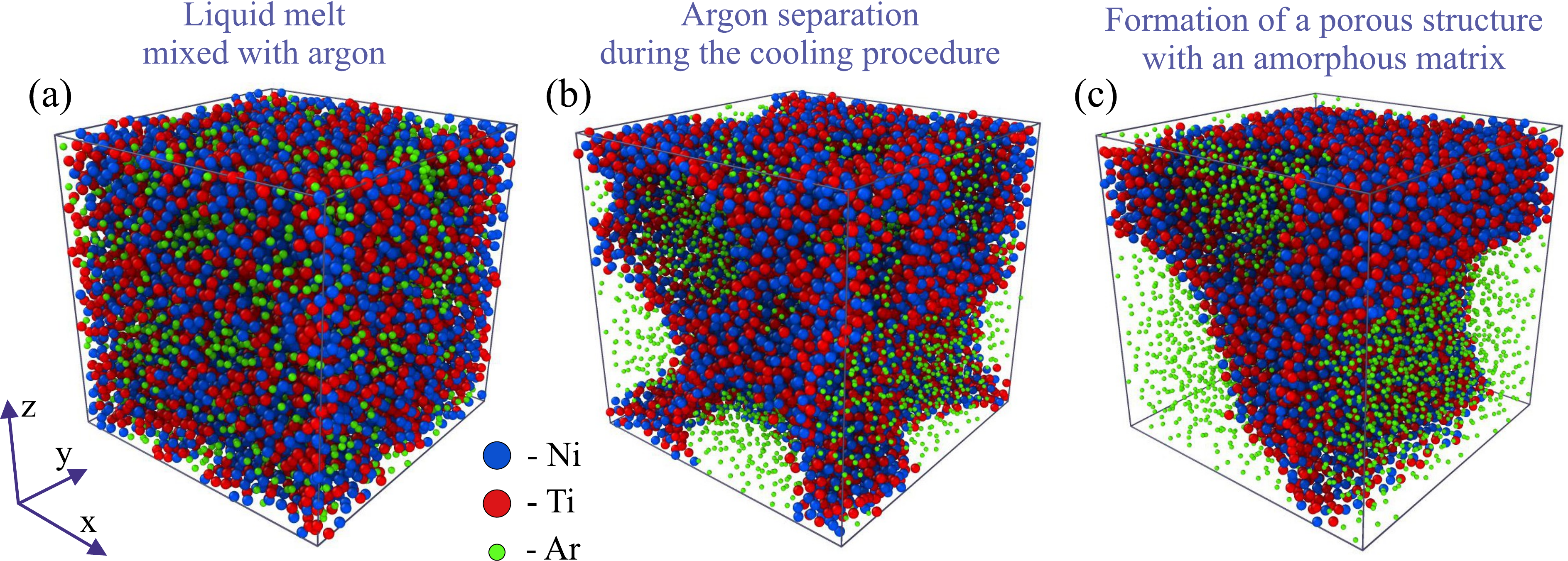}
	\caption{Snapshots of the system in various states: (a) liquid melt mixed with argon; (b) redistributed Ni, Ti and Ar atoms at the stage of cooling of the liquid melt; (c) porous structure and amorphous matrix upon cooling of the liquid melt with injected argon.}
	\label{fig_1}
\end{figure} 

The argon injection procedure and the preparation of porous amorphous nitinol are carried out through the following steps:
\begin{itemize}
	\item The liquid sample is melted at the temperature $T=1.2T_{m}$, where the melting temperature is $T_{m}\simeq1570$\,K, and, then, it is brought to the thermodynamic equilibrium state during the time $0.1$\,ns.
	
	\item The injection of Ar atoms is carried out. This procedure means the random replacement of some Ni and Ti atoms by Ar atoms [see. Figure~\ref{fig_1}(a)]. Here, the proportion of Ni and Ti atoms in the system always remains the same and corresponds to Ni$_{50}$Ti$_{50}$ alloy. The liquid samples are prepared by injection of argon in the fractions $f=10$, $14$, $15$, $16$, $17$, $18$, $19$, $20$ and $23$\% from the total number of atoms in the system. After the injection of argon, the system is again brought to the thermodynamic equilibrium state.
	
	\item The liquid nitinol mixed with argon is rapidly cooled to the temperature $300$\,K at the cooling rate $1\times10^{12}$\,K/s. The redistribution of the components occurs during the cooling procedure, as a result of which argon is separated from amorphous nitinol [see Fig. Figure~\ref{fig_1}(b) and~\ref{fig_1}(c)]. The separation of argon gas and solid nitinol occurs heterogeneously. This leads to the formation of the foamed amorphous samples, where the percolating network of pores filled with argon. Note that the porosity and distribution of pores in the system may be dependent on the cooling rate applied to generate the porous amorphous system~\cite{Nagaumi_2001,Carlson_Beckermann_2007}. 
		
	\item The foamed samples were held at the temperature $300$\,K for the time $0.1$\,ns for stabilization of the amorphous matrix. Then, argon is completely removed from the system. The resulting porous amorphous samples are held again at the temperature $300$\,K.
\end{itemize}
The porosity $\phi$ of the obtained samples is determined by the well-known expression ~\cite{Flint_2002_porosity}:
\begin{equation}
\phi = \left(1 - \frac{\rho}{\rho_0} \right)\cdot100\%,  
\end{equation}
where $\rho$ is the density of a porous sample, $\rho_0=6.21$\,g/cm$^3$ is the density of nitinol without pores at the temperature $T=300$\,K.

To avoid the artifacts due to the molecular dynamics simulations of a porous system, the following condition must be satisfied: the average linear pore size must be less than the linear size of the simulation box. In this case, there will be no significant artifacts generated by the periodic boundary conditions. In the present study, the linear size of the simulated system is set so that it exceeds the characteristic size of the pores formed in the system. For example, in the case of the porosity $25$\%, the linear size of the system is $L\approx14.8$~nm, while the average linear pore size is $l\approx3$~nm. In the case of the maximum considered porosity $70$\%, we have $L\approx26$~nm and $l\approx20$~nm.
	
\section{Porosity as function of Argon fraction}

\begin{figure}[ht!]
	\centering
	\includegraphics[width=0.8\linewidth]{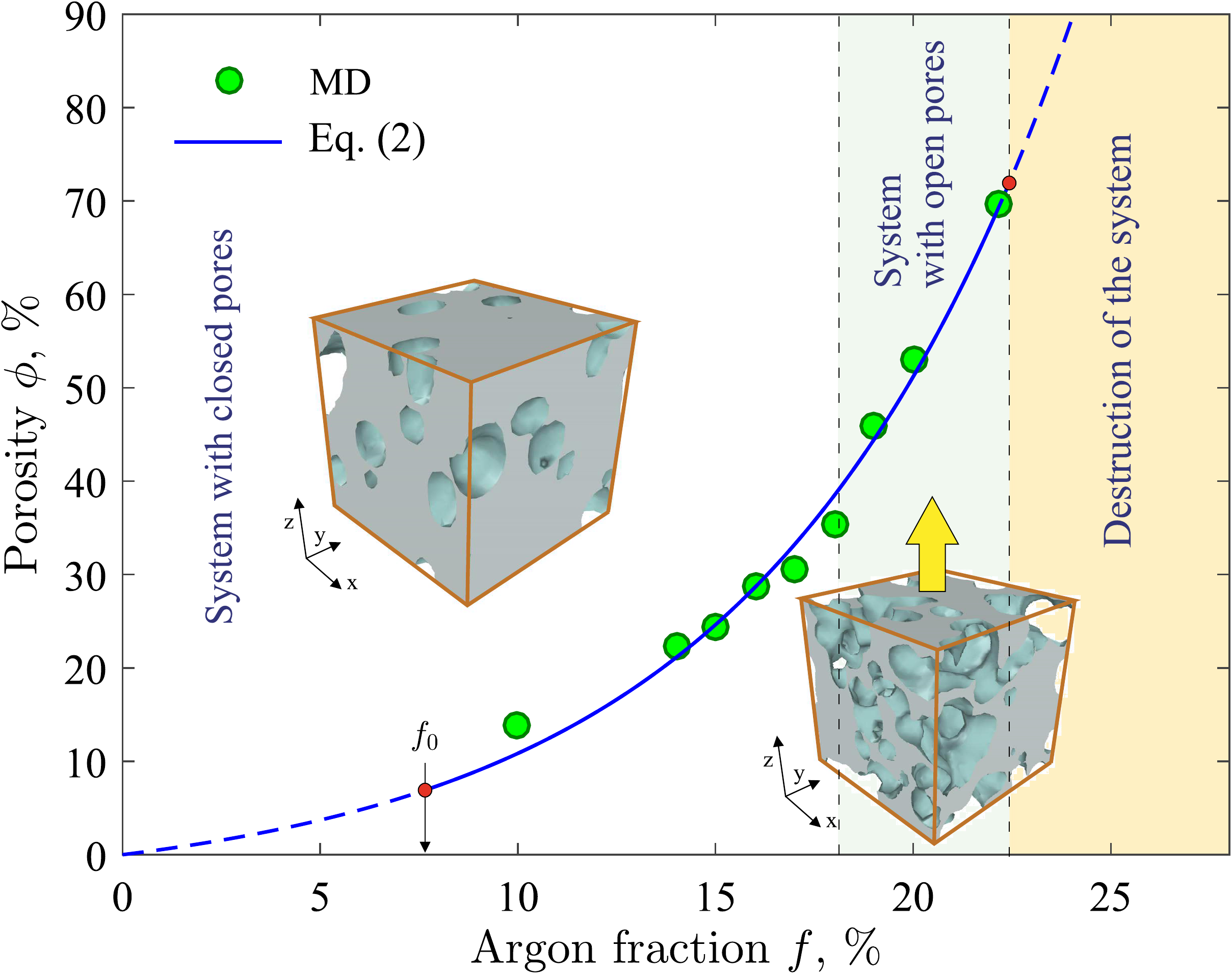}
	\caption{Dependence of the porosity $\phi$ on the injected argon fraction $f$. Regions corresponding to the amorphous system with different porous structures (with open pores, with closed pores) are identified. The system instability region is shown, where the addition of Ar atoms leads to the destruction of the system.}
	\label{fig_2}
\end{figure}

Figure~\ref{fig_2} shows the dependence of the porosity $\phi$ on the injected argon fraction $f$. We find that porous amorphous nitinol contains only closed pores at argon less than $\sim18$\.\%. In this case, the system porosity does not exceed $\phi=35$\%. The resulting pores are stable and have a shape close to spherical. Average linear size of the closed pores is $\sim 2$\,nm, which is a low threshold value for porous alloys belonging to the class of nanoporous materials~\cite{Mishira_2019_size_nanoporous}.
To obtain the porous structure with open pores and the percolating amorphous matrix, the fraction of argon must be above $18$\%. Here, the argon fraction from $f\sim18$\% to $23$\% is optimal. In this case, the porosity of the system varies in the range $\phi\in[40;\,70]$\%. It is important to note that the fraction of the injected argon should not exceed $\sim23$--$25$\,\% at the considered thermodynamic conditions. The structure of the system is destroyed with an increase in the argon fraction above this value: the liquid melt is decayed by overheated argon. Thus, the optimal fraction of the injected argon required to obtain the highly porosity system with open pores is about $\sim20$\% at the considered thermodynamic conditions.
	
We find that the dependence of the system porosity on the injected argon fraction can be reproduced by the exponential law
\begin{equation}\label{eq_porous_fit}
\phi(f) = \phi_0\left[\mathrm{e}^{f/f_0}-1\right],\,\,f\geq f_{0}.
\end{equation}
Here, $\phi_0$ is the fraction of the free volume in the system without pores; this parameter is related to the packing fraction $f_{p}$, $\phi_0\approx1-f_{p}$~\cite{kittel,Filion_Dijkstra_2009}. The parameter $f_{0}$ characterizes the minimum fraction of the injected argon that is sufficient for the formation of stable pores. Expression (\ref{eq_porous_fit}) is an empirical result and it is realized at $f\geq f_0$. In fact, Eq. (\ref{eq_porous_fit}) indicates that the density of the porous system $\rho$ is related to the fraction of injected gas $f$ according to the exponential law
\begin{equation}\label{eq_density_func}
\rho(f) = \rho_0(1+\phi_0-\phi_0\mathrm{e}^{f/f_{0}}),\,\,f\geq f_{0}.
\end{equation}
Both Eqs.~(\ref{eq_porous_fit}) and (\ref{eq_density_func}) take into account the physical effect associated with the fact that there is a minimum fraction of injected gas $f_0$, at which stable pores are formed in the system and the system is not able to ``heal'' these pores. The values of the parameters $f_0$ and $\phi_0$ were determined by fit of Eq.~(\ref{eq_porous_fit}) to the simulation data. For the considered system, the found value $f_0=(7.6\pm0.8)$\% determines the minimum argon fraction that should be injected into nitinol melt to obtain the minimum porosity $\phi(f_0 )\approx1.72\phi_0 \approx 6.9$\%, where $\phi_0=(4.0\pm1.0)$\%. The formed pores will coalesce after removal of Ar atoms from the system if value of the parameter $f_0$ is less than $7.6$\%. We find that for the case minimal porosity with $\phi(f_0)$ it is satisfied the correlation relation $f_0\approx1.9\phi_0$. Then, from expression (\ref{eq_density_func}) at $f=f_0$ we find
\begin{equation}\label{eq_density_limit}
\rho(f_{0})\approx\rho_0(1-0.9f_{0}).
\end{equation}
Expression (\ref{eq_density_limit}) shows the ultimate density of the system, at which the formation of stable pores becomes possible. 

\begin{figure}[ht!]
		\centering
		\includegraphics[width=0.4\linewidth]{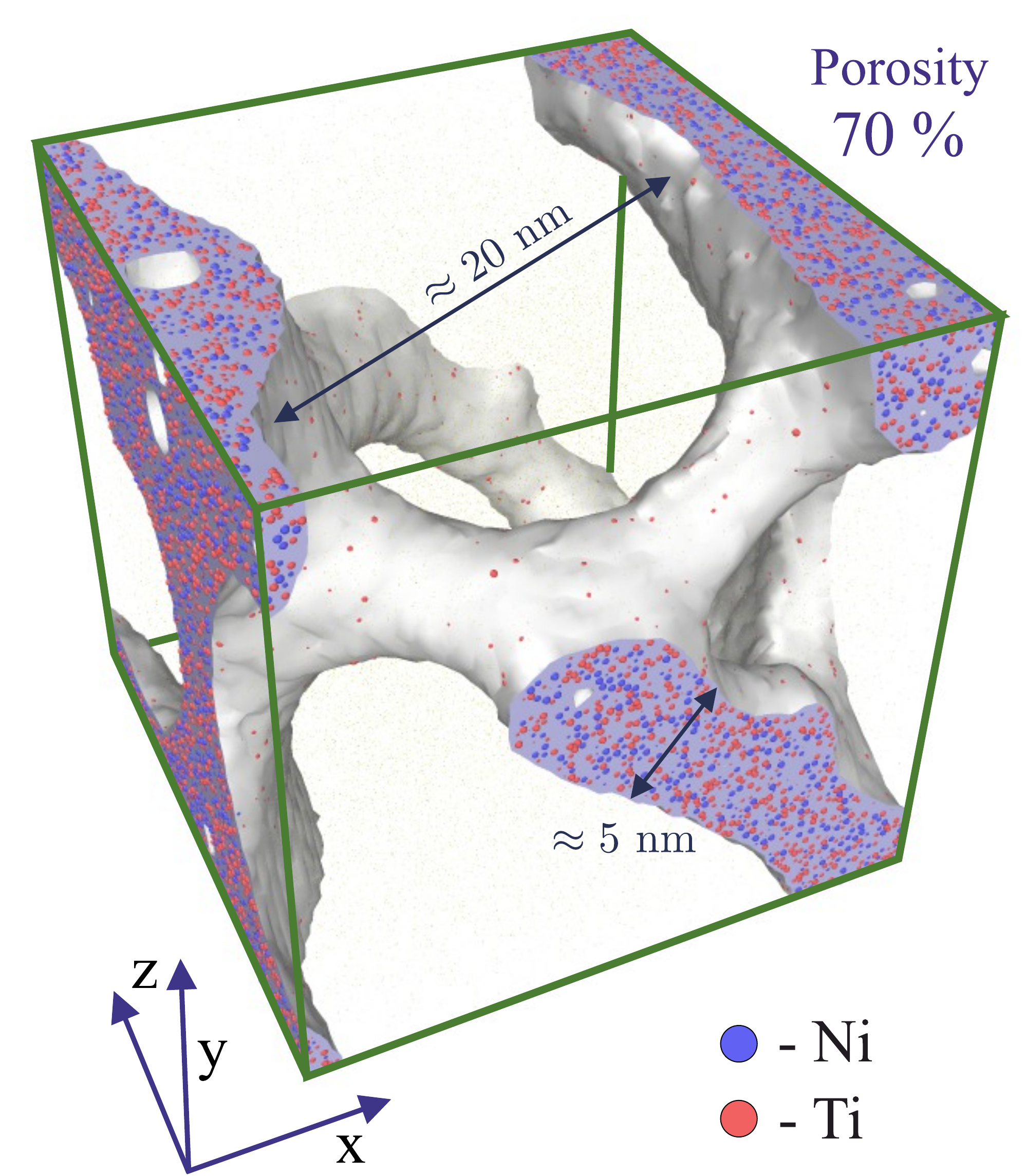}
		\caption{Snapshot of the system with porosity $\sim70$\% prepared by injecting $23$\% argon.}
		\label{fig_3}
\end{figure}

Figure~\ref{fig_3} shows the fragment of the amorphous matrix of porous nitinol with the porosity $70$\%. As can be seen from this figure, the so-called spongy porous structure is formed at injected of more than $20$\% argon. The reason for formation of such spongy structure is the thermal expansion of the liquid melt due to the large fraction of overheated noble gas. In this spongy porous structure the average thickness of the interpore walls is much less than the average linear size of the pores: average thickness of interpore walls is $\sim 5$\,nm, while the average linear size of the pores is $\sim20$\,nm. 
We note that nitinol is in the state of viscous liquid at the considered temperature $\simeq1890$\,K, while argon at this temperature is in strongly overheated gas phase (the melting temperature of solid argon is $\sim88$\,K). Despite the high fraction of argon, the high viscosity of liquid nitinol does not allow the gas phase to be completely separated from the melt. Thus, the partial coalescence of the formed gas bubbles and foaming of liquid nitinol are occured. The high viscosity of the melt also contributes to formation of the percolating matrix at the cooling procedure [see Figure~\ref{fig_3}]. It is noteworthy that such the spongy porous structure almost completely corresponds to the structure of organic foams, aerogels and metallic foams, the porosity of which ones usually is larger than $70$\%~\cite{Bhattacharya_2002_metal_foams,Sehaqui_2011_high_porous_aerogel,Qiu_Yan_2021,Qiu_Du_2021,Qiu_Zou_2018}. Average value of the linear size of the pores in the considered system is also close to the pore size in aerogels, where pores can have sizes from $\sim10$\,nm to $\sim100$\,nm~\cite{Daero_2019_pore_sizes_variety}. Such the spongy porous structure could be of great interest for various practical applications: for example, for the design of filter elements, fuel reservoirs and battery electrodes~\cite{Liu_2014_appl_porous_mats}.

\section{Conclusion}
	
Based on the results of molecular dynamics simulations, we have shown the possibility of obtaining porous amorphous nitinol by injecting argon into the corresponding melt. The energies and forces between Ni, Ti, and Ar atoms were correctly calculated using the hybrid interparticle interaction potential, which made it possible to obtain the porous system whose structure is close to experimentally synthesized aerogels and metallic foams. A relationship between the fraction of the injected argon and the system porosity has been determined, which is well reproduced by an exponential function. We have determined the optimal fraction of the injected argon, which makes it possible to obtain the porous amorphous nitinol with the spongy structure and with the stable amorphous matrix without crystalline inclusions. The limiting fraction of the injected argon was also found, the excess of which leads to the complete extrusion of argon from the liquid melt and to destroy of the system. It is shown that for formation of stable porous structure, the minimum fraction of the injected argon should not be less than $\sim7.6$\% at the considered thermodynamic conditions. The obtained results show that well-known method of direct melt foaming by gas injection~\cite{Banhart_2001,Brothers_2006} can be adapted for the synthesis of porous amorphous nitinol if the rapid cooling protocol will be implemented.

The results of the present study are practically significant and can be applied to improve the foaming protocols of metal melts. For example, based on the results presented in Figure~\ref{fig_2}, it is possible to determine the optimal ratio of foaming gas and metal melt to obtain a sample with the required porosity and pore morphology. This makes it possible to simplify the procedure for synthesizing samples with desired mechanical properties for specific applications, for example, for the manufacture of construction materials or implants~\cite{Prasad_Bazaka_2017,Aihara_Zider_2019}. Moreover, the ultimate value of the argon fraction necessary to obtain a highly porous system with the porosity $\sim70$\% was determined for the first time. At such porosity, we have a sponge-like porous structure as in the case of aerogels, organic and metal foams. This result is of great importance in production of porous filters and biocompatible porous nanoparticles, for example, for the delivery and storage of drugs in humans~\cite{Malloy_Quintana_2021,Carcer_Baeza_2020,Zhou_Shen_2017}.
	
\section*{Acknowledgments}
\noindent The work was supported by the Russian Science Foundation (project No. 19-12-00022-P). AVM is grateful to the Foundation for the Development of Theoretical Physics and Mathematics ``Basis'', which contains part of the work related to the implementation of theoretical calculations. The review part about properties of perspective materials is supported by the Kazan Federal University Strategic Academic Leadership Program (PRIORITY-2030).


\section*{References}
\begin{thebibliography}{45}

\bibitem{nitinol_main} Kapoor D 2017 Nitinol for Medical Applications: A brief introduction to the properties and processing of nickel titanium shape memory alloys and their use in stents {\it Johnson Matthey Technological Review} {\bf 61} 66--76

\bibitem{review_niti_applictions} Ishizaki K, Komareni S, Nanko M 1998 {\it Applications of porous materials. In: Porous Materials} (Springer, Boston, MA)
	
\bibitem{review_niti2} Chaudhari R, Vora J J, Parikh D M 2020 {\it Recent Advances in Mechanical Infrastructure Proceedings of ICRAM 2020} (Springer)

\bibitem{Bansiddhi_2011} Bansiddhi A, Sargeant T D, Stupp S I, Dunand D C 2011 Porous NiTi for bone implants: a review {\it Acta Biomaterialia} {\bf 4} 773--782

\bibitem{niti_production} Yasenchuk Y et al. 2019 Biocompatibility and clinical application of porous TiNi alloys made by Self-Propagating High-Temperature synthesis (SHS) {\it Materials} {\bf 12} 2405

\bibitem{Augat_2006} Augat P, Schorlemmer S 2006 The role of cortical bone and its microstructure in bone strength {\it Age and ageing} {\bf 35-S2} ii27--ii31

\bibitem{niti_lifetime} Bosingnore C 2017 Present and future approaches to lifetime prediction of superelastic nitinol {\it Theoretical and Applied Fracture Mechanics} {\bf 92} 298--305

\bibitem{nakas2014_fatigue} Nakas G I, Asik E E, Tunca B and Bor S 2014 Fatigue and fracture behavior of porous TiNi alloys ({\it THERMEC 2013 : International conf. on processing and manufacturing of advanced materials : processing, fabrication, properties, applications, December 2-6, 2013, Las Vegas, USA} vol~783)  ed B Mishra, M Ionescu and T Chandra. (Durnten-Zurich : Trans Tech Publications Ltd.)
	
\bibitem{nitinol_amorph_rem} Gunther V et al. 2019 Formation of pores and amorphous-nanocrystalline phases in porous TiNi alloys made by self-propagating high-temperature synthesis (SHS) {\it Advanced Powder Technology} {\bf 4} 673--680
	
\bibitem{GBN_2021} Galimzyanov B N, Mokshin A V 2021 Mechanical response of mesoporous amorphous NiTi alloy to external deformations {\it International Journal of Solids and Structures} {\bf 224} 111047

\bibitem{Panico_2008_test_porous_niti} Panico M and Brinson L C 2008 Computational modeling of porous shape memory alloys {\it International Journal of Solids and Structures} {\bf 45} 5613--5626

\bibitem{niti_isobaric} Galimzyanov B N, Mokshin A V 2020 Amorphous Porous Phase of Nitinol Generated by Ultrafast Isobaric Cooling {\it Solid State Phenomena} {\bf 310} 150--155

\bibitem{Anikeev_2020_2d_powder_niti} Anikeev S, Yakovlev E, Artyukova N, Mamazakirov O, Kaftaranova M and Promakhov V 2020 {\it 2020 7th International Congress on Energy Fluxes and Radiation Effects (EFRE)} (USA: IEEE)

\bibitem{Karaji_2006_3d_print} Karaji J Z, Speirs M, Dadbak S, Kruth J-P, Weinans H, Zadpoor A A and Yavari S A 2016 Additively manufactured and surface biofunctionalized porous nitinol {\it ACS Applied Material Interfaces} {\bf 9} 1293--1304 

\bibitem{Shiskovsky_2012_sls_porous_nitinol} Shiskovski I 2012 Hysteresis modeling of the porous nitinol delivery system, designed and fabricated by SLS method {\it Physics Procedia} {\bf 39} 893--902

\bibitem{Biswas_2005_SHS} Biswas A 2005 Porous NiTi by thermal explosion mode of SHS: processing, mechanism and generation of single phase microstructure {\it Acta Materialia} {\bf 53} 1415--1425

\bibitem{Galimzyanov_Yarullin_2018}
Galimzyanov B N, Yarullin D T and Mokshin A V 2018 Change in the crystallization features of supercooled liquid metal with an increase in the supercooling level {\it JETP Letters} {\bf 107} 629--634

\bibitem{Galimzyanov_Nikiforov_2020} Galimzyanov B N, Nikiforov G A and Mokshin A V 2020 Effect of Ultrafast Cooling on Pore Formation in Amorphous Titanium Nickelide {\it Acta Physica Polonica A} {\bf 137} 1149--1152

\bibitem{JIANG_2013_amorph_cryst} Jiang S, Tang M, Zhao Y, Hu L, Zhang Y and Liang Y 2013 Crystallization of amorphous NiTi shape memory alloy fabricated by severe plastic deformation {\it Transactions of Nonferrous Metals Society of China} {\bf 24} 1758

\bibitem{Galimzyanov_Mokshin_2018} Galimzyanov B N and Mokshin A V 2018 Morphology of critically sized crystalline nuclei at shear-induced crystal nucleation in amorphous solid {\it Journal of Rheology} {\bf 62} 265--275
	
\bibitem{Banhart_2001} Banhart J 2001 Manufacture, characterisation and application of cellular metals and metal foams {\it Progress in Materials Science} {\bf 46} 559--632

\bibitem{Brothers_2006} Brothers A H and Dunand D C 2006 Amorphous metal foams {\it Scripta Materialia} {\bf 54} 513--520

\bibitem{lammps} Thompson A P et al 2022 LAMMPS -- a flexible simulation tool for particle-based materials modeling at the atomic, meso, and continuum scales {\it Comp. Phys. Comm.} {\bf 271} 108171

\bibitem{niti_pot} Ko W S, Grabowski B and Neugebauer J 2015 Development and application of a Ni-Ti interatomic potential with high predictive accuracy of the martensitic phase transition {\it Phys. Rev. B} {\bf 92} 134107

\bibitem{Lee_Baskes_2000}
Lee B-J, Baskes M I 2000 Second nearest-neighbor modified embedded-atom-method potential {\it Phys. Rev. B} {\bf 62} 8564

\bibitem{lj_proof} Daun K J, Titantah J T and Karttunen M 2012 Molecular dynamics simulation of thermal accommodation coefficients for laser-induced incandescence sizing of nickel particles {\it Applied Physics B} {\bf 107} 221--228

\bibitem{Hansen_1969_lj_pot} Hansen J-P and Verlet L 1969 Phase transitions of the Lennard-Jones system {\it Physical Review} {\bf 184} 151--161

\bibitem{Nagaumi_2001}
Nagaumi H 2001 Prediction of porosity contents and examination of porosity formation in Al–4.4\%Mg DC slab {\it Science and Technology of Advanced Materials} {\bf 2} 49--57

\bibitem{Carlson_Beckermann_2007}
Carlson K D, Lin Z, Beckermann, C 2007 Modeling the Effect of Finite-Rate Hydrogen Diffusion on Porosity Formation in Aluminum Alloys {\it Metall. Mater. Trans. B} {\bf 38} 541--555	

\bibitem{Flint_2002_porosity} Flint E L and Flint A L 2002 {\it 2.3 Porosity. In: Methods of soil analysis, Part 4: Physical Methods} (Soil Science Society of America, Inc.)

\bibitem{Mishira_2019_size_nanoporous} Mishra R, Militky J and Venkataraman M 2019 {\it 7 -- nanoporous materials. In: Nanotechnology in textiles: theory and application} (Woodhead Publishing)

\bibitem{kittel} 
Kittel Ch 2005 {\it Introduction to Solid State Physics} (John Wiley \& Sons, Inc.)

\bibitem{Filion_Dijkstra_2009}
Filion L, Dijkstra M 2009 Prediction of binary hard-sphere crystal structures {\it Phys. Rev. E} {\bf 79} 046714

\bibitem{Bhattacharya_2002_metal_foams} Bhattacharya A, Calmidi V V and Mahajan R L 2002 Thermophysical properties of high porosity metal foams {\it International Journal of Heat and Mass Transfer} {\bf 45} 1017--1031
	
\bibitem{Sehaqui_2011_high_porous_aerogel} Sehaqui H, Zhou Q and Berglund A L 2011 High-porosity aerogels of high specific surface area prepared from nanofibrillated cellulose (NFC) {\it Composites Science and Technology} {\bf 71} 1593--1599

\bibitem{Qiu_Yan_2021}
Qiu L, Yan K, Feng Y, Liu X, Zhang X 2021 Bionic hierarchical porous aluminum nitride ceramic composite phase change material with excellent heat transfer and storage performance {\it Composites Communications} {\bf 27} 100892

\bibitem{Qiu_Du_2021}
Qiu L, Du Y, Bai Y, Feng Y, Zhang Z, Wu J, Wang X, Xu C 2021 Experimental Characterization and Model Verification of Thermal Conductivity from Mesoporous to Macroporous SiOC Ceramics {\it J. Therm. Sci.} {\bf 30} 465--476

\bibitem{Qiu_Zou_2018}
Qiu L, Zou H, Tang D, Wen D, Feng Y, Zhang Z 2018 Inhomogeneity in pore size appreciably lowering thermal conductivity for porous thermal insulators {\it Applied Thermal Engineering} {\bf 130} 1004--1011
	
\bibitem{Daero_2019_pore_sizes_variety} Daero L, Jinyoung K, Seohyun K, Gunhwi K, Jihun R, Sangrae L and Haksoo H 2019 Tunable pore size and porosity of spherical polyimide aerogel by introducing swelling method based on spherulitic formation mechanism {\it Microporous and Mesoporous materials} {\bf 288} 109546

\bibitem{Liu_2014_appl_porous_mats} Liu P S and Chen G F 2014 {\it Application of Porous Metals. In book: Porous Metals} (Tsinghua University Press Limited, Elseiver)

\bibitem{Prasad_Bazaka_2017}
Prasad K, Bazaka O, Chua M, Rochford M, Fedrick L, Spoor J, Symes R, Tieppo M, Collins C, Cao A, Markwell D, Ostrikov K, Bazaka K 2017 Metallic Biomaterials: Current Challenges and Opportunities {\it Materials} {\bf 10} 884

\bibitem{Aihara_Zider_2019}
Aihara H, Zider J, Fanton G, Duerig T 2019 Combustion Synthesis Porous Nitinol for Biomedical Applications {\it International Journal of Biomaterials} {\bf 2019} 4307461

\bibitem{Malloy_Quintana_2021}
Malloy J, Quintana A, Jensen C J, Liu K 2021 Efficient and Robust Metallic Nanowire Foams for Deep Submicrometer Particulate Filtration {\it Nano Lett.} {\bf 21} 2968--2974

\bibitem{Carcer_Baeza_2020}
Parra-Nieto J, Cid M, Carcer I, Baeza A 2020 Inorganic porous nanoparticles for drug delivery in antitumoral therapy {\it Biotechnol. J.} {\bf 16} 2000150

\bibitem{Zhou_Shen_2017}
Zhou M, Shen L, Lin X, Hong Y, Feng Y 2017 Design and pharmaceutical applications of porous particles {\it RSC Adv.} {\bf 7} 39490--39501

\end{thebibliography}
\end{document}